# How Do Users Interact with an Error-prone In-air Gesture Recognizer?

**Ahmed Sabbir Arif[1], Wolfgang Stuerzlinger[1], Euclides Jose de Mendonca Filho[2], Alec Gordynski[3]**

[1]York University
Toronto, Ontario, Canada
{asarif, wolfgang}@cse.yorku.ca

[2]Federal University of Bahia
Salvador, Bahia, Brazil
euclidesmendonca.f@gmail.com

[3]Flowton Technologies
Toronto, Ontario, Canada
ag@flowton.ca

## ABSTRACT
We present results of two pilot studies that investigated human error behaviours with an error prone in-air gesture recognizer. During the studies, users performed a small set of simple in-air gestures. In the first study, these gestures were abstract. The second study associated concrete tasks with each gesture. Interestingly, the error patterns observed in the two studies were substantially different.

**ACM Classification Keywords**
H.5.m User Interfaces: Miscellaneous.

## INTRODUCTION
Invisible user interfaces are becoming increasingly popular. It refers to an interface that is either invisible or becomes invisible with successive learned interactions. Users interact with such interfaces mainly via gestures, potentially in the air [11]. Smart televisions, video game consoles, or the Leap Motion enable such interactions. Current gesture recognizers achieve up to 99% accuracy [1], provided that sufficient training data is available and reliable input technologies are used. However, in practice, gesture-based techniques are more error prone than traditional ones, likely due to gesture ambiguity [8] and the lack of appropriate feedback [5]. As the number of easily *performable* gestures is limited, most techniques utilize the same or similar gestures for multiple tasks [8]. This makes it difficult for users to associate gestures to concrete tasks and for the system to disambiguate them. The absence of direct visual feedback for in-air gestures poses additional problems [11].

A well-regarded theory in psychology error research, called the mismatch concept [3], holds both the user and system responsible for committing errors, but attributes errors to the mismatch in the interaction among these two. This implies that a deeper insight into how users interact with error prone systems is crucial for user-friendly interfaces and effective recognition techniques. For instance, if a large number of users experience a given mistake, it is prudent to globally change that specific system feature. Similarly, an adaptive system could calibrate itself to individual user behaviours. Yet, error behaviours for gesture interfaces have not been well studied. To inform in-air gesture research, we conducted two pilot studies to explore error behaviours.

## INVESTIGATED GESTURES
Five simple in-air gestures, *push*, *left*, *right*, *up*, and *down*, were used in the studies. These gestures were performed by placing the dominant hand in the *rest* position (resembling a stop or wait gesture), moving the hand *forward*, *left*, *right*, *up*, or *down*, respectively, and then bringing it back to the rest position. Illustrated in Figure 1. The rest position was used as the origin and was updated during each *push* gesture.

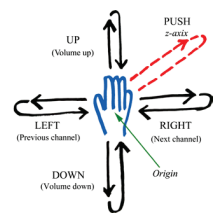

**Figure 1. The five gestures used during the pilot studies.**

### Performance Metrics
The following metrics were calculated during the pilots.

*Gesture Attempts (GA):* The average number of attempts it took to perform a gesture. A flawless gesture recognizer will result in a GA of one, provided there was no human error.

*Error Rate (ER):* This is the average percentage of errors committed with the system. This is a compound of the *Human Error Rate ($ER_H$)* and the *System Error Rate ($ER_S$)*. The first is the average committed by the users, with the second committed by the system. The experimenter manually recorded all user actions and system reactions. Incidents where users performed the correct gesture, but the system failed to (correctly) recognize it, were classified as system errors. If users performed the wrong gesture, this was seen as a human error. The experimenter also kept a watchful eye for incidents where both the human and the system made mistakes. However, such incidents did not occur during the pilots.

## PILOT STUDY 1
This study investigated abstract gestures, i.e. gestures that were not associated with concrete tasks.

### Apparatus and Participants
A custom application, developed with the OpenNI [10], was used during the pilot. It sensed motion using a Microsoft

Kinect and recognized gestures using the NiTE™ computer vision library [9].

Fourteen participants, aged from 20 to 35 years, average 23.6, voluntarily participated in this pilot. All were right-handed. Seven were male. Three frequently interacted via in-air gestures with Kinect, Wii, and/or PS3 controllers, nine occasionally, and the rest had no prior experience.

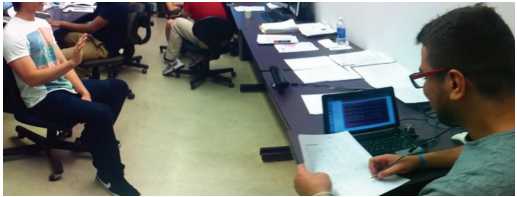

Figure 2. Pilot study 1 setup. The Kinect was placed on a table 20″ above the floor. The participant sat approximately 40″ away from the Kinect facing the device and inputted gestures as instructed by the experimenter (bottom right). Here, the participant is holding his hand in the initial *rest* position. The experimenter manually recorded the inputted gestures, failed attempts, and the types of errors.

## Procedure and Design

Participants were instructed to place their hands in the initial *rest* position and to wait for the experimenter's instructions. The experimenter then guided them through the study by instructing them on the gestures to perform. The experimenter also verbally informed them of any mistakes made by them or by the system. In other words, the experimenter simulated auditory feedback for the system. Figure 2 illustrates the study setup. If an error was made, participants were requested to try the same gesture again until it was successfully recognized by the system. If no error was made, they were asked to input the next gesture. The experimenter manually recorded the type and numbers of attempted gestures, successful or unsuccessful attempts, and the types of errors committed. Upon completion of the study, participants were interviewed on the perceived easiness of the gestures, fatigue, and general comments. We uses a within-subjects design: 14 participants × 4 blocks × 4 trials × 6 gestures per trial (*push*, *left*, *right*, *up*, *down*, *push*) = 1,344 gestures in total. The four inner gestures were counterbalanced in each block via a balanced Latin square. The first and the last *push* gesture activated and deactivated the gesture recognition system, correspondingly. There was no practice, but the experimenter demonstrated how to perform the gestures before the study.

## Results

An Anderson-Darling test revealed that the study data was normally distributed. A Mauchly's test confirmed that the data's covariance matrix was circular in form. Therefore, we used repeated-measure ANOVA for all analysis.

### Gesture Attempts (GA)

An ANOVA failed to identify a significant effect of gesture type on GA ($F_{4,13} = 0.13$, ns). There was also no significant effect of block ($F_{3,13} = 1.62$, $p > .05$) or gesture type × block ($F_{12,156} = 1.08$, $p > .05$). On average, the overall GA during the four blocks were 1.08 (SE = 0.02), 1.09 (SE = 0.02), 1.05 (SE = 0.02), and 1.03 (SE = 0.01), respectively. Figure 3 illustrates the average GA for each gesture.

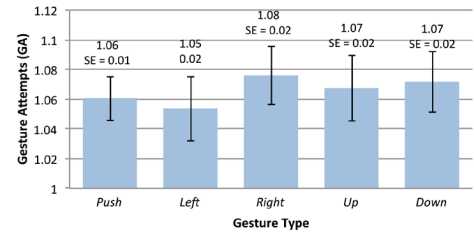

Figure 3. Average gesture attempts (GA) for the five gestures. The error bars represent ±1 standard error (SE).

### Error Rate (ER)

An ANOVA failed to identify a significant effect of gesture type on ER ($F_{4,13} = 0.49$, ns). There was also no significant effect of block ($F_{3,13} = 1.62$, $p > .05$) or gesture type × block ($F_{12,156} = 1.01$, $p > .05$). On average, overall ER during the four blocks were 7.44 (SE = 2.2), 8.33 (SE = 2.38), 5.06 (SE = 1.53), and 3.27% (SE = 1.05), respectively. Figure 4 illustrates the average ER for each gesture.

Experimenter records revealed that 80.3% of all errors were committed by the system. The remaining 19.7% were human errors. A Chi-squared test found this to be statistically significant ($\chi^2 = 36.00$, $df = 1$, $p < .0001$).

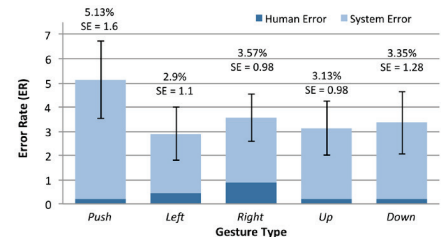

Figure 4. Average error rate (ER) for the five gestures. The error bars represent ±1 standard error (SE).

An ANOVA failed to identify a significant effect of gesture type on $ER_H$ ($F_{4,13} = 1.07$, $p > .05$). There was also no significant effect of block ($F_{3,13} = 1.42$, $p > .05$) or gesture type × block ($F_{12,156} = 1.05$, $p >.05$). Likewise, an ANOVA failed to identify a significant effect of gesture type on $ER_S$ ($F_{4,13} = 0.62$, ns). There was no significant effect of block ($F_{3,13} = 2.28$, $p > .05$) or gesture type × block ($F_{12,156} = 0.81$, ns).

### Human Effort and Fatigue

No straightforward trend was observed regarding this. About half the users reported high post-study fatigue, while the rest were mostly neutral. A Chi-squared test did not find this to be statistically insignificant ($\chi^2 = 0.999$, $df = 2$, ns).

## Discussion

The system recognized 94% of all gestures correctly. As at least 97% accuracy rate is necessary for the users to find a gesture-based system useful [7], our setup served well as an *error prone system*. The results showed that gesture type has no significant effect on attempts per gesture or accuracy. This means none of the gestures were substantially more error prone than the others. One potential explanation is that the gestures were abstract and users did not have to associate tasks with each one of them. There was also no significant effect of block. This is not unusual considering the length of the study. More interestingly, we identified the following behaviours by observation and user responses.

- About 57% users found the *down* gesture uncomfortable to perform. One potential reason is that they were performing this gesture in a seated position, which did not give them enough space to freely move their hands (far enough) downwards. Other gestures were rated mostly neutral.
- About half the users made relatively shorter gestures by the second block, i.e. they kept their hands closer to their body compared to the first block.
- Users often briefly got confused between *left* and *right*. That is, they started performing the opposite gesture, i.e. *left* instead of *right* or vice versa, but corrected themselves almost immediately. The experimenter did not record such incidents as errors.

**PILOT STUDY 2**

This pilot study investigated task-associated gestures, i.e. gestures that were associated with concrete tasks.

**Apparatus and Participants**

A custom application, developed with OpenNI [10], was used during this study. It sensed motion using a Microsoft Kinect and recognized gestures using the NiTE™ computer vision library [9]. The application was shown full-screen on a 21″ CRT monitor, to mimic a television screen. It displayed the current channel number in digits at the centre, and current volume level in a slider at the bottom of the screen. The monitor also served as a stand for the Kinect. The application logged all interactions with timestamps. The experimenter used a separate application to present random gestures in each session and to record user behaviours and errors. It was launched on a laptop computer to facilitate fast logging via keyboard shortcuts, see Figure 5.

Seven participants, aged from 21 to 43 years, average 29.0, voluntarily participated in this pilot. They all were right-handed and two were female. One frequently interacted via in-air gestures with a Nintendo Wii or Leap Motion and two occasionally, while the rest had no prior experiences. They all received a small compensation.

**Procedure and Design**

Participants were instructed to place their hands in the initial *rest* position and to make a *push* gesture when they were ready. This started a session and presented a random task on the top of the screen for them to perform. Participants performed four tasks: load the previous channel (*left*), load the next channel (*right*), raise the volume (*up*), and lower the volume (*down*), see Figure 1. Error correction was forced. That is, participants had to keep trying until the system performed the intended task. They were provided with visual feedback—they could see the channel changing and the volume bar moving upon successful recognition of the corresponding gestures. Upon completion of the study, they were asked to fill out a short questionnaire. A within-subjects design was used: 7 participants × 3 sessions × 64 gestures (*left*, *right*, *up*, *down*, each 16 times, *randomized*) = 1,344 gestures in total. There was no practice block, but the experimenter demonstrated how to perform the tasks prior to the pilot. There was a mandatory 5 min. break between sessions.

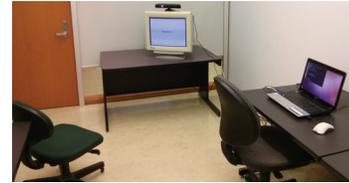

**Figure 5. Pilot study 2 setup. Participants sat on the left chair facing the monitor and the Kinect. The experimenter sat on the right chair with a clear view of the setup and the participant. The chairs were approximately 18″ high. The Kinect was placed above the monitor in 4° angle and approximately 46″ above the floor. The distance between the participants and the Kinect was kept at approximately 48″. The experimenter used an extended keyboard, not visible here, to log user behaviours.**

**Results**

An Anderson-Darling test revealed that the study data was normally distributed. A Mauchly's test confirmed that the data's covariance matrix was circular in form. Therefore, repeated-measure ANOVA was used for all analysis.

*Gesture Attempts (GA)*

An ANOVA identified a significant effect of gesture type on GA ($F_{3,6} = 5.7$, $p < 0.01$). A Tukey-Kramer test revealed that *left* took significantly more attempts than *up* and *down*. However, no significant effect of session ($F_{2,6} = 0.47$, ns) or gesture type × session ($F_{6,36} = 0.37$, ns) was identified. On average, the overall GA during the three sessions were 1.07 (SE = 0.02), 1.05 (SE = 0.01), and 1.1 (SE = 0.02), respectively.

Figure 6 illustrates the average GA for each gesture.

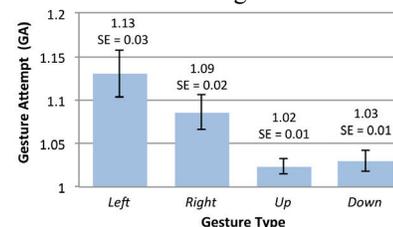

**Figure 6. Average gesture attempts (GA) for the four gestures. The error bars represent ±1 standard error (SE).**

*Error Rate (ER)*

An ANOVA identified a significant effect of gesture type on ER ($F_{3,6} = 5.87$, $p < .01$). A Tukey-Kramer test revealed that *left* suffered from significantly more errors than *up* or *down*. Yet, no significant effect of session ($F_{2,6} = 0.48$, ns) or gesture type × session ($F_{6,36} = 0.41$, ns) was found. The overall average ER during the three sessions were 6.70 (SE = 1.91), 4.69 (SE = 1.0), and 8.93% (SE = 4.30). Figure 7 illustrates the average ER for each gesture.

Experimenter records revealed that 91.2% of all errors were committed by the system and the remaining 8.8% by humans. A Chi-squared test found this to be statistically significant ($\chi^2 = 67.24$, $df = 1$, $p < .0001$).

An ANOVA found no significant effect of gesture type on $ER_H$ ($F_{3,6} = 0.26$, ns). There was also no significant effect of session ($F_{2,6} = 0.18$, ns) or gesture type × session ($F_{6,36} = 1.25$, $p > .05$). Yet, an ANOVA identified a significant effect of gesture type on $ER_S$ ($F_{3,6} = 6.48$, $p < .005$). A Tukey-Kramer test revealed that *left* had significantly more system errors than *up* and *down*. There was no significant effect of session ($F_{2,6} = 0.37$, ns) or gesture type × session ($F_{6,36} = 0.36$, ns).

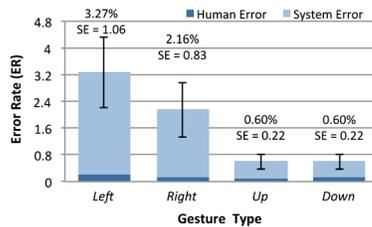

**Figure 7. Average error rate (ER) for the four gestures. The error bars represent ±1 standard error (SE).**

*Human Effort*

A Friedman test failed to identify significance with respect to mental ($\chi^2 = 2.467$, $df = 3$, ns), physical ($\chi^2 = 0.864$, $df = 3$, ns), or temporal demand ($\chi^2 = 1.429$, $df = 3$, ns), performance ($\chi^2 = 6.414$, $df = 3$, ns), effort ($X^2_{(3)} = 1.333$, $df = 3$, ns), or frustration ($\chi^2 = 4.761$, $df = 3$, ns) for the four gestures. Also, no straightforward trend was found regarding fatigue. About 43% users reported that they experienced high post-study fatigue, while the rest were mostly neutral.

## DISCUSSION

The system recognized about 93% of all gestures, which is comparable to the first pilot. However, unlike the first pilot, significant effects of gesture type were observed for both gesture attempts and accuracy. This indicates the possibility that error patterns may be different for meaningless and meaningful gestures. One explanation is that recall-based tasks are more challenging. Remarkably, the *left* gesture was significantly more error prone and took more attempts than the other ones. The reason is that users often performed the *right* or the *down* gestures instead of *left*. While we do not have a definite reason for this behaviour, the fact that all our participants were right-handed may have contributed to this phenomenon. Similar to the first pilot, there was no significant effect of session. User behaviours towards the system were also comparable to the first pilot. But this time we collected more data to further analyse the behaviours.

- About 71% users believed that their performance got faster with time. The results do not support this. An ANOVA failed to identify a significant effect of session on task completion time ($F_{2,6} = 1.7$, $p > .05$). It is possible that users' gesture recall time decreased with practice. However, this is difficult to verify in a short study as the difference between the novice and the expert recall and preparation time is only ~600ms [6].

- Users easily got impatient with system errors and made repetitive attempts without allowing the system to react to the first re-attempt. This caused additional system errors (which we recorded correspondingly). About 60% errors took more than one attempt to fix. Compared to behaviours in a different error prone system [2], a Chi-squared test found this to be statistically significant ($\chi^2 = 4.0$, $df = 1$, $p < 0.05$). During the interviews, all users stated that their reaction was instinctive.

- About 43% users experienced high post-study fatigue. Observation revealed that these users continued making longer gestures (moved their hands further away from their body), while the others started making shorter gestures by the second session.

- Users often got confused between the *left* and the *right* gestures. About 82% of the total errors were committed while performing these two gestures.

## CONCLUSION AND FUTURE WORK

We presented results of two pilot studies that investigated error behaviours with an error prone in-air gesture recognizer. In the first study users performed abstract gestures, while in the second they performed task-associated gestures. Results show that although error patterns during the two studies were substantially different, users reactions to the errors were similar.

The first pilot instructed users verbally to perform a task and provided them with auditory feedback, while the second pilot used a custom application and provided visual feedback. In the future, we will investigate the effect of these factors and of different user expertise on the study results.